\documentclass{article}
\usepackage{spconf,amsmath,amssymb,caption,graphicx,hyperref,booktabs,multirow,enumitem,float}
\usepackage[numbers]{natbib}

\title{Auditory-Feedback-Conditioned Flow TTS for Implicit Lombard Adaptation}
\twoauthors
 {Yunya Wang}
	{School of Data Science\\
	The Chinese University of Hong Kong, Shenzhen\\
    yunyawang@link.cuhk.edu.cn}   
 {Satoshi Nakamura\sthanks{\textnormal{Corresponding Author}}\hspace{2em}}
	{\hspace{-3em}School of AI\\
	\hspace{-3em}The Chinese University of Hong Kong, Shenzhen\\
    \hspace{-3em}snakamura@cuhk.edu.cn}
\begin{document}
%
\maketitle
\begin{abstract}
Conventional text-to-speech (TTS) systems can lose intelligibility in noise environments because they cannot adapt to environmental changes. However, humans tend to increase their vocal efforts in noisy environments, a phenomenon known as the Lombard effect. Building on an auditory-feedback Lombard TTS, we investigate whether a pretrained Flow-based TTS can learn acoustic adaptation directly from auditory feedback without explicit prosodic control. Signal-to-noise ratio (SNR) feedback and automatic speech recognition (ASR) feedback are extracted from the noise-corrupted waveforms and injected into the Flow-matching condition for utterance refinement and causal block-level updates. We assess feedback benefits across prompting and noise conditions using plain/Lombard prompting, prompt-switching controls, and paired-finetuning. Experiments show that while finetuning provides strong robustness under in-domain conditions, auditory feedback yields additional cue-dependent benefits that generalize to unseen noise. Moreover, stronger power/F0 tracking alone does not guarantee better speech intelligibility or quality.
\end{abstract}
\section{Introduction}
\label{sec:intro}

Humans maintain high speech intelligibility by monitoring their own speech during production. When the environment gets noisy, they increase vocal effort, intensity, and pitch, a phenomenon known as the Lombard effect \cite{lane1971lombard}. Although conventional TTS systems generate high-quality speech in clean conditions, their intelligibility can degrade in noise. Early studies on Lombard speech synthesis relied on signal processing techniques, such as incremental synthesis for online updates \cite{rottschafer2015online}, or hidden Markov model (HMM)-based acoustic modulation \cite{raitio2011analysis}. More recent works used transfer learning to induce Lombard speaking style \cite{paul2020enhancing,bollepalli2019lombard}. Other works use explicit conditioning or predefined acoustic embeddings \cite{akti2026synthesizing, woszczyk2025voice}. However, these methods do not integrate a ``perception-generation" loop for adaptation to dynamic noise.

The machine speech chain introduced a human-like mechanism of ``speaking while listening'' by jointly training ASR and TTS \cite{tjandra2017listening}. Building on this framework, prior work demonstrated the feasibility of Lombard speech synthesis driven by SNR, ASR, and power feedback under static and dynamic noise \cite{novitasari2023self,novitasari2022machine}. However, it relied on  autoregressive architectures with explicit prosodic control. It remains unclear whether modern TTS models can infer the appropriate acoustic response directly from auditory feedback without explicit controls of pitch, intensity or other acoustic features. 

Flow-based TTS models have demonstrated a strong capability in synthesizing natural and expressive speech \cite{woszczyk2025voice,chen2025f5}. To investigate the above question, we inject environmental and recognition-oriented feedback directly into the acoustic condition of a pretrained LLM-Flow-HiFT TTS architecture, allowing feedback-conditioned adaptation without explicit acoustic control. Our main contributions are summarized as follows:

\begin{itemize}[
    leftmargin=1.2em,
    itemsep=0pt,
    topsep=2pt,
    parsep=0pt
]
    \item We investigate auditory feedback conditioning in a pretrained LLM-Flow-HiFT TTS architecture, allowing acoustic adaptation without explicit pitch, power or prosodic controls. 
    \item We evaluate auditory feedback under plain and Lombard prompts, using prompt-switching controls and a paired-finetuning baseline to examine its acoustic responses.
    \item We characterize cue-specific adaptation through power/F0 tracking and noise-stage transitions under dynamic noise.
\end{itemize}

\begin{figure}[t]
    \centering
    \includegraphics[
        width=\columnwidth,
        trim=0 10 0 0,
        clip
    ]{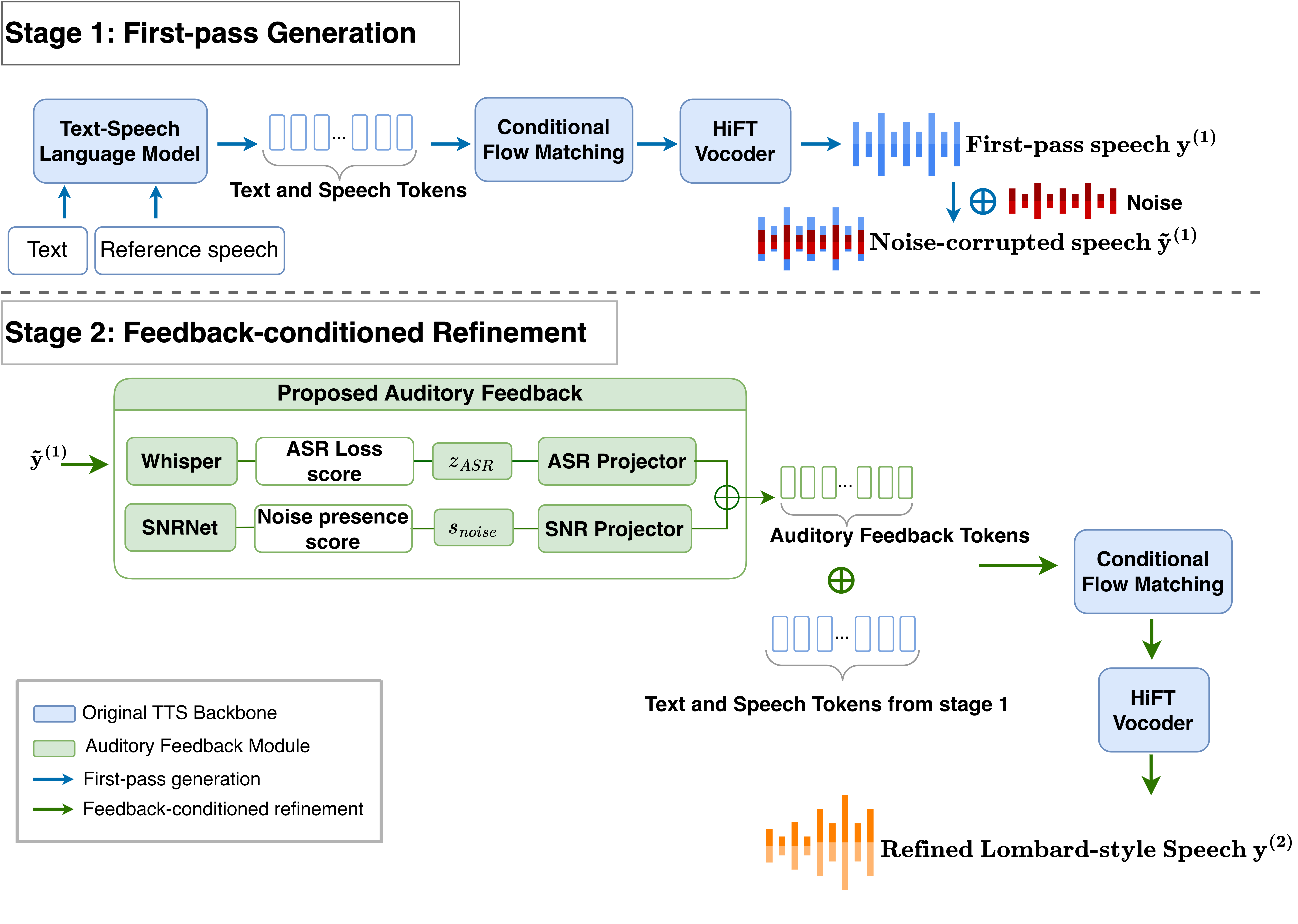}
    \captionsetup{skip=2.0pt}
    \caption{Auditory-feedback-conditioned TTS framework. Static inference refines a complete first-pass utterance using SNR/ASR feedback extracted from its noise-corrupted waveform. Dynamic inference applies the same feedback causally. Each block uses the latest valid feedback condition for generation.}
    \label{fig:main_figure}
\end{figure}

\section{Methods}
\label{sec:methods}

\subsection{Overview}
\label{sec:overview}
We build on CosyVoice2, which consists of a text-to-speech language model (LM), a conditional Flow-matching acoustic model, and a HiFT neural vocoder \cite{du2024cosyvoice}. As shown in Fig \ref{fig:main_figure}, we augment the Flow condition with two auditory feedback representations extracted from noise-corrupted speech $\tilde{\mathbf{y}}$. We introduce an environmental signal
$s_{\mathrm{noise}}$ and a recognition-oriented representation
$z_{\mathrm{ASR}}$. Both are projected into the Flow conditioning space for speech refinement.

\subsection{SNR Feedback} 
\label{subsec:snr-feedback}
Given the log-Mel representation
$X$ of the observed waveform $\tilde{y}$, a pretrained SNRNet produces posterior probabilities over three noise levels: clean, 0 dB, and -10 dB. 
\begin{equation}
\mathbf p =
\mathrm{softmax}(W_{\mathrm{cls}}f_{\mathrm{SNRNet}}(X)+b_{\mathrm{cls}}).
\end{equation}

Then, we convert these probabilities into a continuous noise-presence score:
\begin{equation}
s_{\mathrm{noise}}
=1-p_{\mathrm{clean}}
=p_{0\mathrm{dB}}+p_{-10\mathrm{dB}},
\end{equation}
The score represents the model's confidence that the observed speech is noise-corrupted.

\subsection{ASR Feedback} 
ASR feedback represents recognition difficulty with a frozen Whisper model. Given noise-corrupted waveform $\tilde{\mathbf y}$ resampled to 16kHz, and reference tokens
$\mathbf{x}=(x_1,\ldots,x_S)$, we compute teacher-forced token-level
cross-entropy sequences:
\begin{equation}
\ell_s=-\log p(x_s\mid x_{<s},\tilde{\mathbf y}),
\qquad s=1,\ldots,S.
\label{eq:asr_loss}
\end{equation}
A temporal convolutional network with pooling encodes the normalized loss: 
\begin{equation}
z_{\mathrm{ASR}} =
E_{\mathrm{ASR}}\!\left(
\mathrm{Norm}([\ell_1,\ldots,\ell_S])
\right).
\label{eq:asr_embed}
\end{equation}

For static refinement, ASR feedback is computed over the complete first-pass utterance. Dynamic inference follows CosyVoice2's default streaming configuration with 25-token blocks. At each block, Whisper transcripts and word timestamps from the cumulative speech are aligned with the reference text. Recognized words ending at least 250 ms before the current audio boundary are eligible for alignment. Match coverage is the proportion of eligible recognized words matched to reference words. The committed reference prefix advances when coverage reaches at least 55\%, and matched words remain stable across two consecutive updates. The corresponding words end-times should differ by less than 0.4\,s. Feedback is extracted from noisy speech over a rolling three-block window, containing at least 3 committed words and 0.8\,s of speech. Otherwise, the previous valid representation is retained. Clean speech is only used for word-boundary estimation, and the noisy speech is used to extract ASR feedback representation for subsequent blocks. 

With batch size one on an NVIDIA A800, a complete ASR feedback update, including Whisper recognition and alignment, takes 214.3/298.9\,ms (median/P95). 

\subsection{Feedback-Conditioned Flow Model}
\label{sec:flow}
Let $h_{\mathrm{tok}}$ denote the frame-level token condition produced by the
CosyVoice2 Flow encoder and length regulator. We incorporate  $s_{\mathrm{noise}}$
and $z_{\mathrm{ASR}}$ through
separate learnable projections:
\begin{equation}
\mu =
h_{\mathrm{tok}}
+ \mathbf{\alpha_{\mathrm{SNR}}}P_{\mathrm{SNR}}(s_{\mathrm{noise}})
+ \mathbf{\alpha_{\mathrm{ASR}}}P_{\mathrm{ASR}}(z_{\mathrm{ASR}}), 
\label{eq:cfm_representation}
\end{equation}
Here, $P_{\mathrm{SNR}}$ and $P_{\mathrm{ASR}}$ project feedback representations
into the Flow conditioning space. The binary gates $\mathbf{\alpha_{\mathrm{ASR}}}, \mathbf{\alpha_{\mathrm{SNR}}}\in \{0,1\} $ select SNR, ASR, or joint SNR+ASR feedback configurations. No F0, intensity or other acoustic attribute is explicitly prescribed, and speaker information follows the original CosyVoice2 conditioning path.

Static refinement uses first-pass feedback. Dynamic generation uses the latest valid feedback from preceding observations. We preserve the original masked conditional Flow-matching objective \cite{du2024cosyvoice}. Auditory feedback modifies only the conditioning of the Flow velocity field. In particular, ASR feedback serves as a conditioning representation rather than an auxiliary loss, avoiding
unstable gradient propagation through the vocoder-ASR cascade.

\section{Experiments}
\label{sec:exp}

\subsection{Dataset}
\label{sec:data}
We train on the Audio-Visual Lombard Grid corpus
\cite{alghamdi2018corpus}, which contains paired plain and Lombard speech from 54 speakers. We use a speaker-disjoint train/validation/test split of 45/2/7 speakers. During training, noisy observations are corrupted with Gaussian white noise at 0/$-10$dB. Clean and noisy observations are paired with plain and Lombard targets, respectively. For evaluation, synthesis texts are drawn from the Wall Street Journal corpus \cite{paul1992design} and Hurricane dataset \cite{cooke2013hurricane}.
\vspace{-10pt}

\subsection{Training Details}
\label{sec:training}
We initialize from the pretrained CosyVoice2-0.5B Flow checkpoint and jointly train the Flow backbone and the auditory feedback modules. We keep the LM, HiFT vocoder, SNRNet and Whisper frozen. $E_{\mathrm{ASR}}$
uses three 1-D convolutional layers and pooling to produce a 512-D representation. $P_{\mathrm{SNR}}$ and
$P_{\mathrm{ASR}}$ project feedback representations to the 80-D Flow condition. Models are trained for 30 epochs using Adam ($1\times10^{-5}$) with the original CosyVoice2 Flow-matching objective. The joint SNR+ASR model is initialized from the SNR-conditioned model and jointly optimized with the ASR feedback for additional 30 epochs. The finetune model uses the same paired data for 30 epochs without feedback modules.

\subsection{Metrics}
\label{sec:metrics}
\textbf{Intelligibility and quality.}
We evaluate noise-corrupted synthesized speech using word error rate (WER) from NVIDIA Parakeet TDT 0.6B v2 \cite{nvidia2025parakeet}, and Short-Time Objective Intelligibility (STOI) between clean speech and its noise-corrupted version. UTMOS \cite{saeki2022utmos} is computed on clean synthesized speech. 

\textbf{Subjective perception.}
We conducted MOS listening tests with 63 human listeners. Each listener rated a subset of 40 static and 32 dynamic clean synthesized clips on a 1-5 scale. Static rating assessed naturalness, and dynamic rating assessed how appropriately the speech adapted to the time-varying noise. Confidence intervals (CIs) use listener-cluster bootstrap.

\textbf{Acoustic adaptivity.} We compute utterance-level lagged Spearman correlations  $\rho_P=\rho(P_t,n_{t-1})$ and \ $\rho_{F0}=\rho(F0_t,n_{t-1})$, where $n$ denotes noise severity of the block (clean $<$ 0 dB $<$ $-10$ dB). For each utterance, $\Delta P$ and
$\Delta F0$ measure absolute differences between the 2nd blocks of adjacent noise stages. Results are aggregated across utterances. Correlations are reported for incremental and recovery trajectories. The abrupt trajectory is characterized by boundary changes only.

\vspace{-10pt}
\subsection{Experimental Settings} 
\label{sec:exp_settings}
We evaluate baseline, SNR, ASR, and joint SNR+ASR systems under plain/Lombard prompts. Static evaluation uses 321 WSJ utterances under clean conditions and Gaussian noise at 10, 0, and $-10$ dB. Dynamic evaluation uses 50 utterances under incremental ($\mathrm{clean}\!\rightarrow\!0\rightarrow\!-10$), recovery ($\!-10\!\rightarrow\!0\rightarrow\mathrm{clean}$) and abrupt ($\mathrm{clean}\!\rightarrow\!-10\rightarrow\!0$) trajectories. Generalization is
evaluated at unseen SNRs (+5/$-5$/$-15$ dB) and unseen dynamic WHAM noise (+5$\rightarrow-5\rightarrow-15$ dB) \cite{wichern2019wham}.

Oracle-prompt and causal-prompt controls switch between plain and Lombard prompts using current and causally delayed noise observations, respectively. These systems aim to examine whether prompt selection alone can reproduce dynamic adaptation. 

For each system, noise is scaled relative to the power of its synthesized speech to achieve the specified SNR. The noise scaling is performed independently for every block. If a noisy block would exceed the noise limit, the complete speech–noise mixture is uniformly peak-scaled to 0.99 after mixing, preserving its specified SNR.

Results marked with ``$*$" indicate a significant difference from the prompt-matched finetune system ( $^*p_{\rm adj}<0.05$ paired permutation test with Holm correction)

\section{Results}
\begin{table}[h]
\captionsetup{skip=1pt}
\caption{Speech intelligibility under static noise conditions.}
\centering
\label{tab:static_wer_stoi}
\small
\setlength{\tabcolsep}{3.5pt}
\fontsize{6.5}{7.1}\selectfont
\resizebox{\columnwidth}{!}{\begin{tabular}{lccccccc}
\toprule
\multirow{2}{*}{System}
& \multicolumn{4}{c}{WER $\downarrow$}
& \multicolumn{3}{c}{STOI $\uparrow$} \\
\cmidrule(lr){2-5}
\cmidrule(lr){6-8}
& Clean & 10 dB & 0 dB & $-10$ dB
& 10 dB & 0 dB & $-10$ dB \\
\midrule

\multicolumn{8}{c}{\textit{Plain prompt}} \\
\midrule
Baseline
& 0.110 & 0.151 & 0.190 & 0.650
& 0.807 & 0.666 & 0.526 \\

Finetune
& 0.094  & 0.122 & 0.177  & 0.591  
& 0.825 & 0.678 & 0.527 \\

+ SNR + ASR
& 0.100 & \textbf{0.110}  & \textbf{0.173} & \textbf{0.557}
& 0.820\rlap{$^{*}$} & 0.681  & 0.526 \\

+ ASR
& 0.091 & 0.111  & 0.183 & 0.580
& \textbf{0.857}\rlap{$^{*}$}  & \textbf{0.701}\rlap{$^{*}$}  & 0.538\rlap{$^{*}$}  \\

+ SNR
& \textbf{0.089} & 0.120 & 0.181 & 0.581 
& 0.817\rlap{$^{*}$}  & 0.690\rlap{$^{*}$}  & \textbf{0.544}\rlap{$^{*}$}  \\

\midrule
\multicolumn{8}{c}{\textit{Lombard prompt}} \\
\midrule
Baseline
& 0.063 & 0.107 & \textbf{0.117} & 0.348
& 0.883 & 0.736 & 0.566 \\

Finetune
& 0.068 & 0.103 & 0.118 & 0.338
& 0.876 & 0.727 & 0.556 \\

+ SNR + ASR
& \textbf{0.057} & 0.093 & 0.130 & 0.349
& 0.875\rlap{$^{*}$} & 0.725\rlap{$^{*}$} & 0.550\rlap{$^{*}$} \\

+ ASR
& 0.080 & 0.099 & 0.156\rlap{$^{*}$} & 0.360
& \textbf{0.898}\rlap{$^{*}$}  & \textbf{0.749}\rlap{$^{*}$}  & \textbf{0.576}\rlap{$^{*}$}  \\
+ SNR
& 0.060 & \textbf{0.093} & 0.122 & \textbf{0.318}
& 0.880 & 0.735 & 0.561\rlap{$^{*}$} \\

\bottomrule
\end{tabular}}
\end{table}
\vspace{-10pt}

\begin{table}[h]
\centering
\caption{Intelligibility and objective speech quality under dynamic noise.}
\label{tab:dynamic_quality}
\scriptsize
\setlength{\tabcolsep}{2.00pt}
\renewcommand{\arraystretch}{0.96}

\resizebox{\columnwidth}{!}{
\begin{tabular}{lccc|@{\hspace{3.5pt}}ccc|@{\hspace{3.5pt}}ccc}
\toprule
& \multicolumn{3}{c}{\textbf{Incremental}}
& \multicolumn{3}{c}{\textbf{Recovery}}
& \multicolumn{3}{c}{\textbf{Abrupt}} \\
\cmidrule(lr){2-4}
\cmidrule(lr){5-7}
\cmidrule(lr){8-10}

System
& WER$\downarrow$ & STOI$\uparrow$ & UTMOS$\uparrow$ 
& WER$\downarrow$ & STOI$\uparrow$ & UTMOS$\uparrow$ 
& WER$\downarrow$ & STOI$\uparrow$ & UTMOS$\uparrow$\\
\midrule

\multicolumn{10}{c}{\textit{Plain prompt}}\\
\midrule

Baseline
& 0.369 & 0.694 & 2.965
& 0.333 & 0.783 & 2.920
& 0.320 & 0.723 & 2.949\\

Finetune
& \textbf{0.282} & 0.697 & \textbf{3.526}
& \textbf{0.207} & \textbf{0.842} & \textbf{3.498}
& 0.239 & 0.738 & \textbf{3.549} \\

+ SNR+ASR
& 0.300 & 0.700 & 3.303\rlap{$^{*}$} & 0.258 & 0.792\rlap{$^{*}$} & 3.246\rlap{$^{*}$}
& \textbf{0.222} & 0.737 & 3.288\rlap{$^{*}$}\\

+ ASR
& 0.354 & \textbf{0.703} & 3.173\rlap{$^{*}$}
& 0.262 & 0.800\rlap{$^{*}$} & 3.134\rlap{$^{*}$}
& 0.276 & 0.735 & 3.177\rlap{$^{*}$}\\

+ SNR
& 0.327 & 0.702 & 3.270\rlap{$^{*}$}
& 0.340\rlap{$^{*}$} & 0.793\rlap{$^{*}$} & 3.246\rlap{$^{*}$}
& 0.277 & 0.734 & 3.266\rlap{$^{*}$}\\

\midrule
\multicolumn{10}{c}{\textit{Lombard prompt}}\\
\midrule

Baseline
& 0.151 & 0.732 & 3.694
& 0.161 & 0.830 & 3.670
& 0.122 & 0.779 & 3.699\\

Finetune
& 0.163 & \textbf{0.718} & \textbf{3.877}
& \textbf{0.133} & \textbf{0.871} & \textbf{3.915}
& \textbf{0.104} & 0.784 & \textbf{3.872} \\

+ SNR+ASR
& \textbf{0.129} & 0.737\rlap{$^{*}$} & 3.687\rlap{$^{*}$}
& 0.148 & 0.826\rlap{$^{*}$} & 3.659\rlap{$^{*}$}
& 0.117 & \textbf{0.785} & 3.648\rlap{$^{*}$}\\

+ ASR
& 0.156 & 0.728\rlap{$^{*}$} & 3.675\rlap{$^{*}$}
& 0.188 & 0.830\rlap{$^{*}$} & 3.688\rlap{$^{*}$}
& 0.129 & 0.775 & 3.643\rlap{$^{*}$}\\

+ SNR
& 0.176 & 0.732 & 3.708\rlap{$^{*}$}
& 0.138 & 0.830\rlap{$^{*}$} & 3.703\rlap{$^{*}$}
& 0.113 & 0.782 & 3.711\rlap{$^{*}$}\\

\midrule
\multicolumn{10}{c}{\textit{Prompt-switching controls}}\\
\midrule

Oracle-prompt
& 0.287 & 0.696 & 3.293
& 0.142 & 0.823 & 3.605
& 0.235 & 0.736 & 3.304\\

Causal-prompt
& 0.324 & 0.695 & 3.249
& 0.272 & 0.799 & 3.214
& 0.239 & 0.737 & 3.240\\

\bottomrule
\end{tabular}
}
\vspace{1pt}
\end{table}

\begin{table}[h]
\centering
\caption{Generalization under unseen static Gaussian noise and dynamic WHAM noise with plain prompt.}
\label{tab:generalization}
\scriptsize
\setlength{\tabcolsep}{1.6pt}
\renewcommand{\arraystretch}{0.96}
\begin{tabular}{lcccccc}
\toprule
\multicolumn{7}{l}{\textit{A. Static: unseen Gaussian SNRs}}\\
\midrule
& \multicolumn{2}{c}{+5 dB}
& \multicolumn{2}{c}{-5 dB}
& \multicolumn{2}{c}{-15 dB}\\
\cmidrule(lr){2-3}
\cmidrule(lr){4-5}
\cmidrule(lr){6-7}
System
& WER$\downarrow$ & STOI$\uparrow$
& WER$\downarrow$ & STOI$\uparrow$
& WER$\downarrow$ & STOI$\uparrow$\\
\midrule
Finetune
& 0.076 & 0.791
& 0.240 & 0.654
& 0.901 & 0.522\\
SNR+ASR
& \textbf{0.072} & 0.788
& \textbf{0.216}\rlap{$^{*}$} & 0.660
& \textbf{0.881} & 0.527\\
ASR
& 0.088 & \textbf{0.817}\rlap{$^{*}$}
& 0.242 & 0.663\rlap{$^{*}$}
& 0.903 & 0.524\rlap{$^{*}$}\\
SNR
& 0.081 & 0.786\rlap{$^{*}$}
& 0.235 & \textbf{0.666}\rlap{$^{*}$}
& 0.883 & \textbf{0.537}\rlap{$^{*}$}\\
\midrule
\multicolumn{7}{l}{
\textit{B. Dynamic: unseen WHAM, $+5\rightarrow-5\rightarrow-15$ dB}
}\\
\midrule
System
& \multicolumn{1}{c}{WER$\downarrow$}
& \multicolumn{1}{c}{STOI$\uparrow$}
& \multicolumn{1}{c}{$\rho_P\uparrow$}
& \multicolumn{1}{c}{$\rho_{F0}\uparrow$}
& \multicolumn{2}{c}{}\\
\midrule
Finetune
& 0.555 & 0.572 & 0.113 & -0.042
& \multicolumn{2}{c}{}\\
SNR+ASR
& \textbf{0.493}\rlap{$^{*}$} & 0.566 & 0.085 & -0.013
& \multicolumn{2}{c}{}\\
ASR
& 0.554 & \textbf{0.586} & \textbf{0.508} & \textbf{0.150}
& \multicolumn{2}{c}{}\\
SNR
& 0.574 & 0.569 & -0.006 & 0.016
& \multicolumn{2}{c}{}\\
\bottomrule
\end{tabular}
\par\vspace{1pt}
\noindent\parbox{\columnwidth}{%
}
\end{table}
\vspace{-20pt}

\subsection{Effects Under In-domain Noise}
\label{subsec:4.1}
Tables \ref{tab:static_wer_stoi} and \ref{tab:dynamic_quality} show that auditory feedback provides gains dependent on cues and metrics under in-domain noise. In static noise conditions, ASR feedback consistently improves STOI across all tested SNRs under both prompts compared with the finetune. WER improvements are cue-dependent. Joint feedback yields the lowest WER across all noisy SNR under plain prompt, and SNR feedback performs best at 10 and $-10$ dB under Lombard prompt. Under dynamic conditions, the finetune provides substantial robustness, and the benefits of auditory feedback become less consistent. 

\subsection{Generalization Across SNRs and Noise Types}
Under unseen noise, auditory feedback provides clearer gains over finetuning. Joint SNR+ASR feedback significantly reduces WER at $-5$ dB. ASR feedback improves STOI across unseen SNRs, and SNR feedback improves STOI and WER at $-5/-15$ dB conditions (Table \ref{tab:generalization}). Under dynamic WHAM noise, joint feedback achieves the lowest WER(0.493), and ASR feedback obtains the highest STOI(0.586). Combining with Sec \ref{subsec:4.1}, these results suggest that auditory feedback provides gains that extend to unseen noise conditions.






\begin{figure}[t]
    \centering
    \includegraphics[
        width=\columnwidth,
    ]{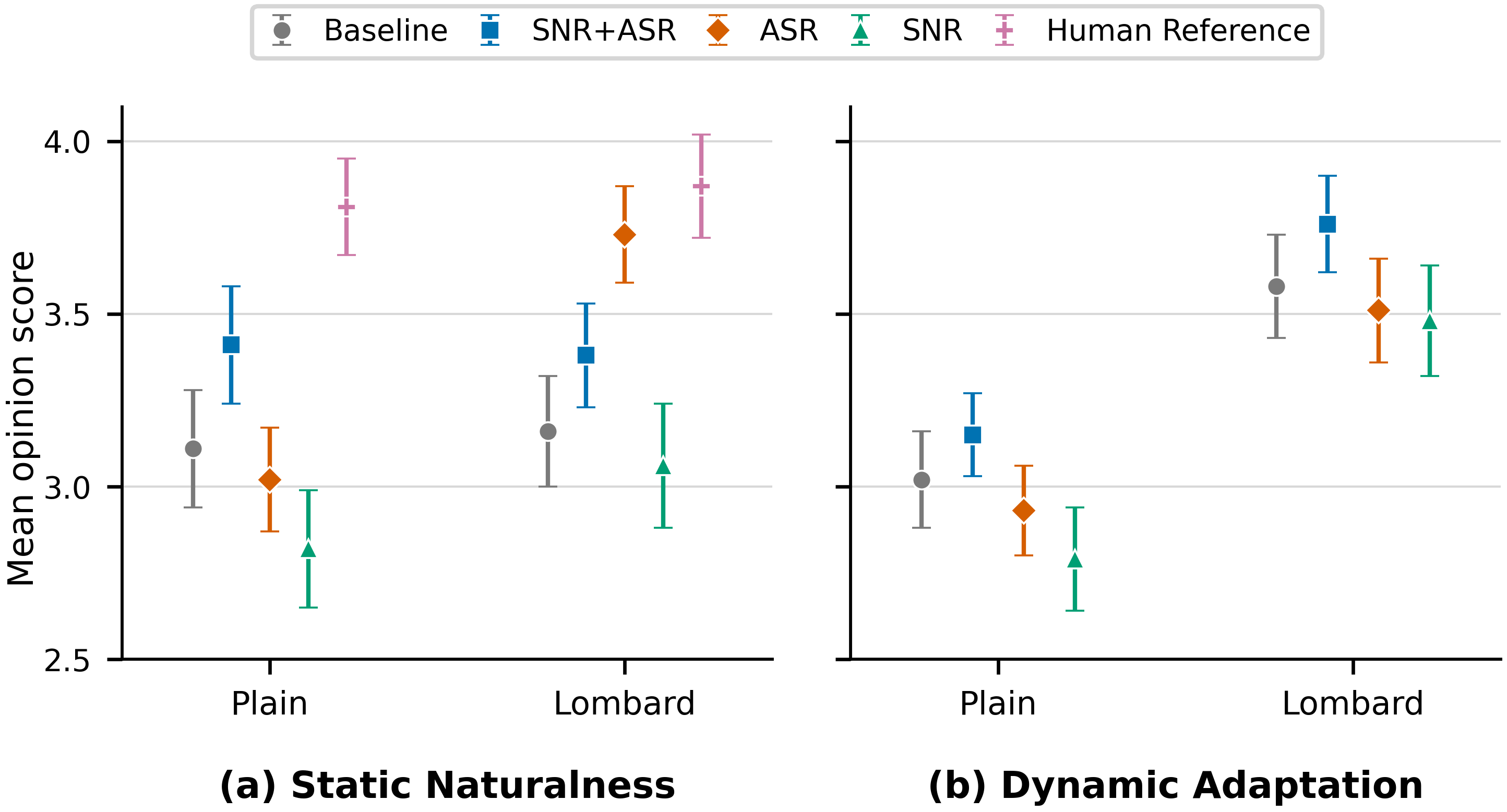}
    \captionsetup{skip=2.5pt,font=small}
    \caption{Subjective MOS results under static and dynamic conditions. Markers show mean ratings, and error bars denote 95\% listener-cluster bootstrap confidence intervals. Static human references use paired plain/Lombard recordings corresponding to evaluation samples.}
    \label{fig:mos_ci}
    \vspace{-10pt}
\end{figure}

\subsection{Subjective Perception}

Figure \ref{fig:mos_ci} shows that joint SNR+ASR feedback yields higher mean MOS ratings than corresponding baselines in all four settings. Meanwhile, ASR feedback shows a pronounced gain in perceptual quality under Lombard-prompting static condition, whereas SNR feedback reduces MOS in all settings. Thus, joint SNR and ASR feedback shows the most consistent perceptual benefits.

\begin{table}[h]
\centering
\caption{Dynamic acoustic adaptation.
Higher $\rho$ values indicate stronger tracking. $\Delta$ denotes absolute between-stage differences in dB/cents.}
\label{tab:dynamic_tracking}

\fontsize{6.5}{7.1}\selectfont
\setlength{\tabcolsep}{2.55pt}
\renewcommand{\arraystretch}{0.96}

\resizebox{\columnwidth}{!}{
\begin{tabular}{
l
cccc|
@{\hspace{3.5pt}}
cccc|
@{\hspace{3.5pt}}
rr
}
\toprule
& \multicolumn{4}{c}{\textbf{Incremental}}
& \multicolumn{4}{c}{\textbf{Recovery}}
& \multicolumn{2}{c}{\textbf{Abrupt}} \\
\cmidrule(l{1pt}r{1pt}){2-5}
\cmidrule(l{1pt}r{1pt}){6-9}
\cmidrule(l{1pt}r{1pt}){10-11}

System
& \multicolumn{1}{c}{\raisebox{0.3ex}{$\rho_P$}}
& \multicolumn{1}{c}{\raisebox{0.3ex}{$\rho_{F_0}$}}
& \multicolumn{1}{c}{$\Delta P$}
& \multicolumn{1}{c|}{$\Delta F_0$}
& \multicolumn{1}{c}{\raisebox{0.3ex}{$\rho_P$}}
& \multicolumn{1}{c}{\raisebox{0.3ex}{$\rho_{F_0}$}}
& \multicolumn{1}{c}{$\Delta P$}
& \multicolumn{1}{c|}{$\Delta F_0$}
& \multicolumn{1}{c}{$\Delta P$}
& \multicolumn{1}{c}{$\Delta F_0$} \\
\midrule

\multicolumn{11}{c}{\textit{Plain prompt}}\\
\midrule

Baseline
& 0.227 & -0.145 & 4.109 & 61.3
& -0.266 & 0.222 & 4.713 & 60.0
& 3.720 & 59.9 \\

Finetune
& 0.108 & -0.036 & 4.965 & 115.0
& 0.130 &  0.056 & 4.987 & 105.0
& 4.821 & 107.5 \\

+SNR+ASR
& 0.392 & -0.171 & 3.860 & 67.5
& \textbf{0.237}\rlap{$^{*}$} & \textbf{0.296} & 4.279 & 77.5
& 4.242 & 62.5 \\

+ASR
& \textbf{0.561}\rlap{$^{*}$} & -0.021 & 3.470 & 78.7
& -0.571\rlap{$^{*}$} & 0.109 & 3.938 & 80.0
& 4.256 & 87.9 \\

+SNR
& 0.534\rlap{$^{*}$} & \textbf{0.289}\rlap{$^{*}$} & 4.286 & 51.3
& 0.199\rlap{$^{*}$} & 0.159 & 4.571 & 50.0
& 4.656\rlap{$^{*}$} & 50.0 \\

\midrule
\multicolumn{11}{c}{\textit{Lombard prompt}}\\
\midrule

Baseline
& -0.012 & -0.160 & 5.570 & 82.5
& 0.061 & 0.142 & 6.052 & 84.4
& 4.844 & 78.4 \\

Finetune
& 0.040 & 0.014 & 5.782 & 102.5
& 0.060 & 0.075 & 4.673 & 94.9
& 6.697 & 90.0 \\

+SNR+ASR
& 0.140 & -0.185 & 4.220 & 87.4
& 0.187 & 0.176 & 5.409 & 82.5
& 4.879 & 78.8 \\

+ASR
& 0.116 & -0.291 & 5.173 & 101.3
& -0.131\rlap{$^{*}$} & \textbf{0.248}
& 4.680\rlap{$^{*}$} & 82.6
& 4.897 & 93.7 \\

+SNR
& \textbf{0.370}\rlap{$^{*}$} & \textbf{0.193}\rlap{$^{*}$} & 4.284 & 84.8
& \textbf{0.346}\rlap{$^{*}$} & 0.199
& 4.204\rlap{$^{*}$} & 65.0
& 4.375 & 72.5 \\

\midrule
\multicolumn{11}{c}{\textit{Prompt-switching controls}}\\
\midrule

Oracle
& 0.679 & 0.426 & 4.079 & 73.8
& 0.516 & 0.425 & 6.222 & 65.0
& 5.027 & 60.0 \\

Causal
& 0.771 & 0.549 & 3.261 & 60.0
& 0.347 & 0.485 & 3.797 & 56.3
& 4.399 & 60.0 \\

\bottomrule
\end{tabular}
}
\end{table}
\vspace{-15pt}

\subsection{Acoustic Adaptation Patterns}
Table \ref{tab:dynamic_tracking} reveals distinct acoustic responses to different feedback cues. SNR feedback exhibits more consistent environment-following behavior, whereas ASR feedback shows more trajectory-dependent power/F0 tracking. Its power correlation changes from 0.561 in incremental trajectory to $-0.571$ in recovery. Also, SNR feedback provides smaller between-stage F0 differences than ASR feedback across all trajectories and prompts, while their power differences depend on condition. 

Furthermore, these acoustic patterns do not directly determine intelligibility or perceived adaptation. Prompt-switching systems yield strong power/F0 tracking without uniformly better intelligibility (Table \ref{tab:dynamic_quality}). Similarly, under WHAM noise, ASR feedback produces better power/F0 tracking but a higher WER than joint feedback (Table \ref{tab:generalization}). Meanwhile, SNR feedback's more consistent tracking is accompanied by lower mean subjective ratings than corresponding baselines (Fig. \ref{fig:mos_ci}). These results suggest that robustness of auditory feedback is not determined by power/F0 tracking alone. 
\vspace{-10pt}

\section{Conclusion}
\label{sec:conclusion}
We investigate auditory feedback conditioning for acoustic adaptation in a pretrained LLM-Flow-HiFT TTS model without explicit prosodic control. While finetuning provides substantial robustness under in-domain conditions, auditory feedback yields additional cue-dependent benefits in unseen noise and induces distinct power/F0 responses. Meanwhile, stronger power/F0 tracking does not consistently yield better intelligibility. These results suggest that modern Flow-based TTS can use auditory feedback for implicit and generalizable acoustic adaptation.

\newpage
\section{Acknowledgment}
The authors would like to acknowledge support from Project W2531054
of the National Natural Science Foundation of China and the Program
for Guangdong Introducing Innovative and Entrepreneurial Teams.

\section{Compliance with Ethical Standards}
This study involved a minimal-risk subjective listening evaluation with volunteer participants. Before participation, all participants were informed of the evaluation task, the intended use of their responses, the voluntary nature of participation, and their right to withdraw. 

\renewcommand{\bibsection}{\section{References}}
\setlength{\bibsep}{1.0ex}       
\bibliographystyle{IEEEbib}
\bibliography{strings,refs}

\end{document}